\documentclass[%
 reprint,
 amsmath,amssymb,
 aps,
]{revtex4-2}

\usepackage{graphicx}
\usepackage{dcolumn}
\usepackage{bm}
\usepackage{xcolor}

\begin{document}

\preprint{APS/123-QED}

\title{Impact of Pauli-blocking effect on optical limiting properties of $WSe_2$ thin films}

\author{Km. Surbhi}
\author{Sourav Bhakta} 
\author{Anupa Kumari}
\author{Utkalika P. Sahoo}
\author{Pratap K. Sahoo}%
\email{pratap.sahoo@niser.ac.in}
\author{Ritwick Das}%
 \email{ritwick.das@niser.ac.in}
\affiliation{%
 School of Physical Sciences, National Institute of Science Education and Research, An OCC of Homi Bhabha National Institute, Jatni, Odisha - 752050, India
}%

\begin{abstract}

We present a detailed investigation on thickness-dependent third-order ($\chi^{(3)}$) nonlinear optical properties of RF-sputtered $WSe_2$ thin-films using ultrashort pulses centered at different excitation wavelengths. The single-beam Z-scan based investigation in the visible spectrum reveals the prominent role of Pauli-blocking towards inducing saturable absorption or optical limiting in $WSe_2$ thin-films of various thicknesses. The study also explores the dispersion in third-order nonlinear susceptibility ($\chi^{(3)}$) in $WSe_2$ thin films. Interestingly, $WSe_2$ thin-films of any thickness exhibit a self-focusing effect depicting a positive nonlinear refractive index ($n_2~>~0$). The frequency-dependent nonlinear absorption ($\beta$) bear distinct correlations with the bandgap of the films which is also investigated through density-functional-theory (DFT) based simulations. The alteration in bandstructure is primarily due to the $Se$-deficiency induced defect bands in $WSe_2$ thin-films which has discernible impact on the optical limiting characteristics. 

\end{abstract}

\maketitle


\section{\label{sec:level1}INTRODUCTION}
Two-dimensional (2D) materials have been the center of research focus in many fields due to their unique and versatile physical, chemical as well as opto-mechanical properties. Graphene, a single layer semi-metallic 2D-carbon architecture having honeycomb lattice geometry, has raised phenomenal interest to investigate its remarkable properties and novel applications \cite{novoselov2005two,geim2010rise,novoselov2012roadmap}. A weak spin-orbit coupling and absence of other interactions in a pristine graphene layer results in a very small, practically non-existent, electronic bandgap. Although, gap opening have been realized through multilayer graphene sheets, doped-graphene layers or by introducing stress, it is appreciably small for a broad range of applications \cite{zhang2009direct,kaloni2014band,hu2018bandgap}. This motivated the search for layered 2D semiconductors (with large molecular weight) which could be grown to a monolayer-scale and they exhibit discernible band opening. Layered transition metal dichalcogenides (TMDs) have received significant attention in this context over the last few years due to their distinctive properties such as high in-plane charge carrier mobility, high mechanical strength and weak coupling between the layers \cite{wilson1969transition,mas20112d,wang2012electronics,butler2013progress,chi20192d}. TMDs have been investigated for a broad range of applications which includes energy storage, gas sensing, biochemical/biological sensing, transistor development, piezoelectric and electronic devices and devising photo-switches \cite{gao2017two,li2012fabrication,hu2019recent,radisavljevic2011single,yang2019coexistence,yin2012single,chi20192d}.\\
TMDs include layered crystal structures with strong (covalent) in plane bonding and weak (vander Waals) out of plane interactions. Recent reports have revealed
that the electronic bandstructure of TMDs is strongly dependent on the number of monolayers and undergoes transition from indirect to direct as a function of thickness \cite{voss1999atomic,yun2012thickness}. Such a possibility endows tremendous functional flexibility from an application point of view. From the perspective of nonlinear optical (NLO) properties, the possibility of tailoring the bandstructure (or bandgap) allows to control the hyperpolarizability and consequently, one could limit the thickness of TMDs in order to suit a certain application. By utilizing the idea, TMD-based architectures have been explored for a broad category of photonic applications such as optical switching \cite{li2016nonlinear,boyd2020nonlinear,jia2020large}, Q-switching and mode-locking \cite{chen2015q,mao2015ws,liu2018tungsten,shao2019wavelength}, data storage \cite{zhou20192d}, optical limiting \cite{loh2006templated,spangler1999recent}, as well as, for optoelectronic applications \cite{ye2021controllable}. A few comprehensive review of NLO characteristics of two-dimensional layered materials could be found in Ref. \cite{autere2018nonlinear,you2019nonlinear}. Conventionally, a single monolayer or a few monolayer TMD has thickness which is about $1000^{th}$-times smaller than the wavelength of light probe. Therefore, the  NLO phase-shifts and absorption by a \emph{few monolayered} TMD-based configuration are relatively weaker. From a practical perspective, TMD thin-films with thicknesses ranging between a hundred nanometer to a few micrometers would be more suitable for photonic applications. However, it is crucial to ascertain whether the properties of a monolayer TMD remain identical (or even similar) for TMD thin-films which is comprised of vertically-stacked multiple monolayers. More importantly, the growth of TMD thin-films could invariably, induce defects which would be essentially dictated by the growth mechanism. Therefore, there remains a strong possibility of modification in the absorption as well as luminescence characteristics of TMD thin-films. From an application viewpoint, it would be very important to investigate the alteration in NLO properties in presence of such defects. In general, TMDs exhibit discernible third-order NLO response owing to small bandgap and consequently, strong hyperpolarizability in the visible/near-infrared spectral domain. This manifests in processes such as third harmonic generation, nonlinear absorption (as saturable absorption (SA), multiphoton absorption) and intensity dependent nonlinear refraction. The dependence of third-order ($\chi^{(3)}$) NLO properties of a few TMD species have been explored with respect to variation in structural parameters and excitation source \cite{wang2014broadband,zhou2015size,zhang2015direct,dong2016dispersion, bikorimana2016nonlinear,wang2019broadband,yan2020third}. However, an investigation aiming at exploring the impact of defects in TMD-based thin-films architecture on $\chi^{(3)}$-induced manifestations is yet to be explored. 

Our work essentially focus on investigating spectral dependence of $\chi^{(3)}$ NLO properties of tungsten-selenide ($WSe_2$) thin-films which have been grown using RF sputtering technique. The NLO investigations, which were carried out using single-beam Z-scan technique, reveal that nonlinear absorption exhibits a saturation behaviour in the visible spectral band owing to \emph{Pauli-blocking} effect. In presence of defects, the impact of \emph{Pauli-blocking} tends to weaken and in the near-infrared (NIR) spectral band, a discernible two-photon absorption (TPA) signature is observed. Consequently, the investigation gives rise to the possibility of deploying RF sputtered $WSe_2$ thin-films for optical limiting and optical switching based opto-electronic applications.  

\section{EXPERIMENTAL DETAILS}
\subsection{Sample preparation}
A set of $WSe_2$ thin films were prepared using a RF magnetron sputtering system with a $WSe_2$ (99.9\% pure) target in an argon environment at room temperature. $WSe_2$  thin films with a desired thickness were sputtered onto a $1~cm \times 1~cm$ $SiO_2/Si$ and glass substrate. Prior to the deposition, the chamber was evacuated by a turbomolecular pump to a vacuum of 3.1 $\times$ $10^{-6}$ mbar. Throughout the  deposition process, the $Ar$ gas concentration was maintained at $15~sccm$, and the power was fixed at $60~W$. The depositions were performed at a pressure of $1\times10^{-2}$ mbar. The number of samples are prepared by varying the time for which deposition takes place which essentially results in a thicker thin film. The morphological features of the RF-sputtered $WSe_2$ thin-films is investigated through field-emission scanning electron microscope (FESEM). The crystal structure and phase analysis is carried out using X-ray diffraction spectrum (from an X-ray diffractometer, Rigaku Smartlab) with Cu K$\alpha$ ($\lambda$ = 1.5418 Å). The Raman spectrum were measured using a Raman spectrometer (Jobin Yvon LabRam  HREvolution, Horiba) with an excitation at $532~nm$ wavelength. The linear absorption spectrum of the $WSe_2$ thin films were carried out using a UV-VIS-NIR spectrometer (Agilent carry 5000 UV-Vis-NIR).
\subsection{Z-scan experimental set up}
 The spectral dependence of NLO characteristics of $WSe_2$ thin-films were investigated through single-beam $Z$-scan technique which used Fourier-transform limited (FTL) ultrashort pulses of temporal width of $\approx~350~fs$ at $515~nm$, $1030~nm$ and $1520~nm$ excitation wavelengths. In order to have discernible optical power at the excitation wavelengths, the pulse repetition rate at $515~nm$ wavelength was chosen to be $1~kHz$ whereas that at $1030~nm$ and $1520~nm$ wavelength was fixed at $100~kHz$ and $504~kHz$ respectively. The optical power of the incident beam is controlled by a combination of half-wave-plate (HWP) and a polarising beam-splitter (PBS). The laser beam is focused to a suitable spot-size using a thin plano-convex lens of focal length $150~mm$. The $WSe_2$ thin films were translated by $\approx~10~cm$ symmetrically about the focal point. An aperture of $1/5^{th}$ the size of laser beam is placed far from the sample for recording the closed-aperture (CA) transmission from the sample. For open-aperture (OA) transmission measurement, the aperture is removed from the beam path. In general, the contribution of OA Z-scan transmission (or nonlinear absorption) is eliminated from the CA transmission measurement by considering a ratio of CA transmission to the OA transmission at a given $z$ coordinate.
\section{RESULTS AND DISCUSSION}
\begin{figure}
\includegraphics[width=0.48\textwidth]{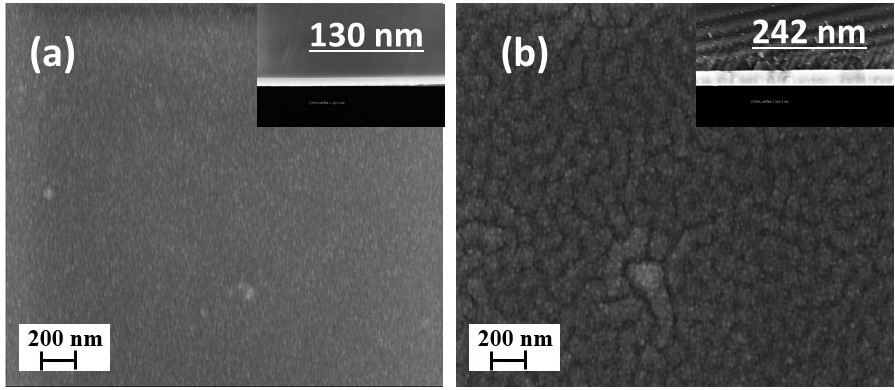}
\caption[Field Emission Scanning Electron Microscopy. ]{Field Emission Scanning Electron Microscopy (FESEM). (a)-(b) shows top surface images of $WSe_2$ thin films having thickness of $130~nm$ and $242~nm$ respectively. Insets show cross-sectional FESEM images for determining thickness.}
\end{figure}
\begin{figure}
\includegraphics[width=0.25\textwidth]{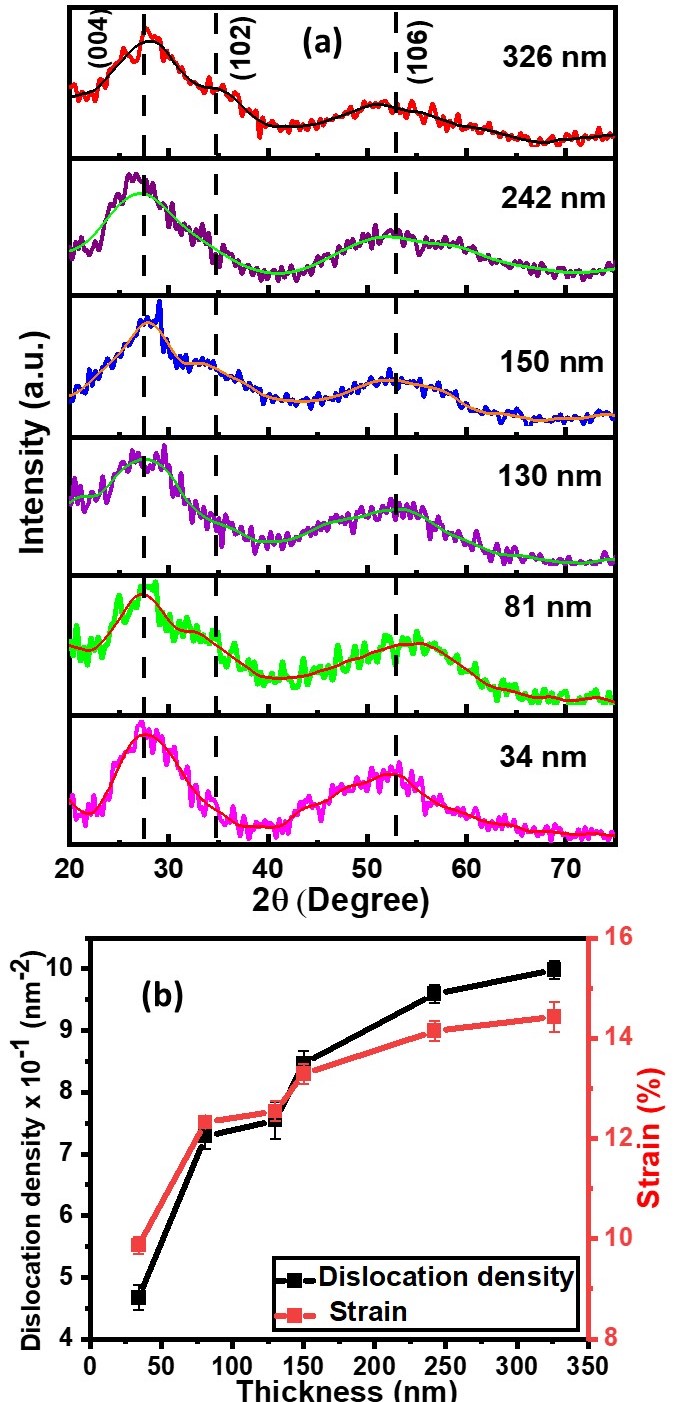}
\caption[X-ray Diffraction pattern.]{(a) X-ray diffraction spectrum of RF-sputtered $WSe_2$ thin films. (b) Variation of strain and dislocation density in $WSe_2$ thin films as a function of thickness.}
\end{figure}
The homogeneity of structure and thickness of the samples is estimated from FESEM images. The cross-sectional FESEM is performed to obtain the thickness of the deposited thin films. The thicknesses were measured to be $34~nm$, $81~nm$, $130~nm$, $150~nm$, $242~nm$ and $326~nm$. In Fig. 1(a) and (b), representative top-view images of $WSe_2$ thin films of thickness $130~nm$ and $242~nm$ respectively are shown. Fig 1(a) appears to exhibit a smooth surface as compared to that shown in Fig. 1(b). The insets represent the respective cross-sectional images. 
\begin{figure}
	\includegraphics[width=0.48\textwidth]{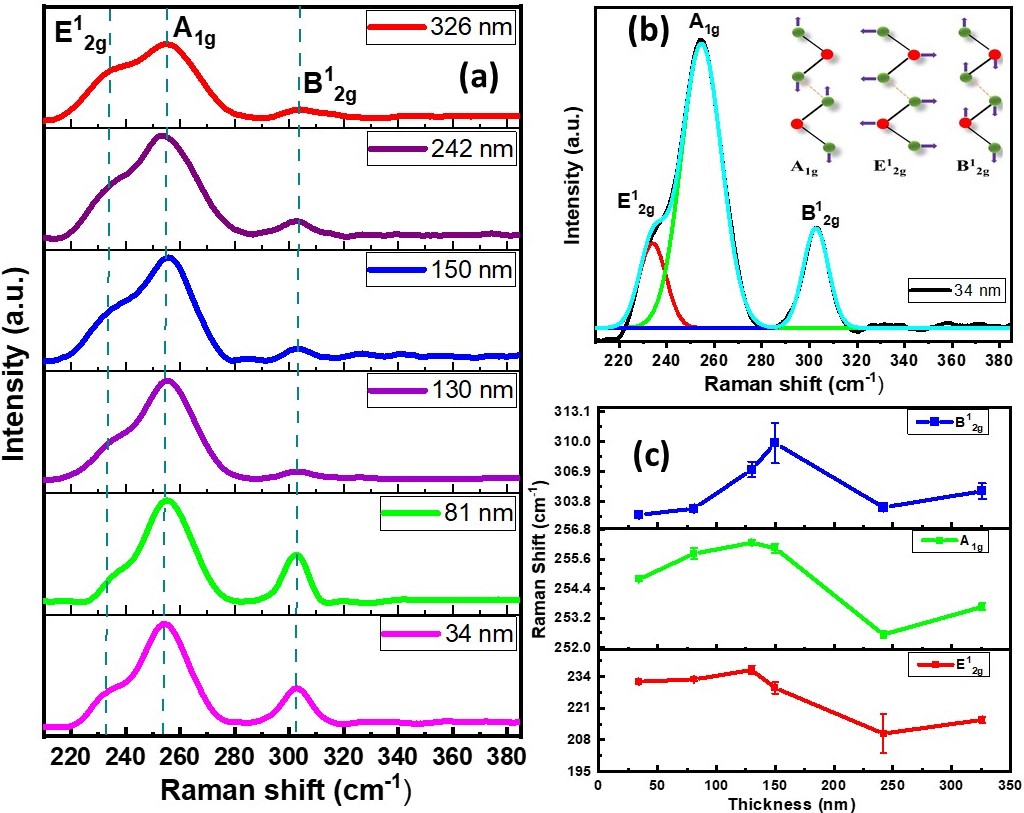}
	\caption[Raman spectrum of the RF-sputtered $WSe_2$ thin films.]{(a) Raman spectrum of $WSe_2$ thin films. (b) Variation of Raman modes of film as a function of film thickness. (c) The multiple-peak fit for $34~nm$ film. Insets show the schematic of possible vibrational modes.}
	\label{fig:Raman}
\end{figure}
\begin{figure}
	\includegraphics[width=0.48\textwidth]{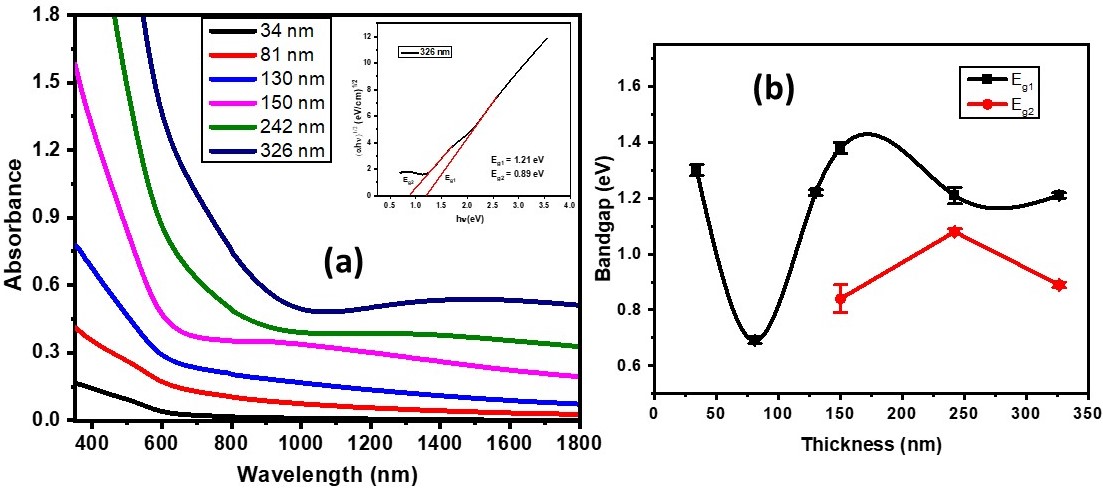}
	\caption[Absorption spectra of $WSe_2$ thin films.]{(a) Absorption spectra of $WSe_2$ thin films. Inset shows the Tauc plot for estimating the bandgap. (b) Variation of bandgap as a function of $WSe_2$ fim thickness.}
	\label{fig:uv-vis}
\end{figure}
The study of crystal structure, crystallite sizes, strain, dislocation density and quantitative phase analysis is carried out using X-ray diffraction (XRD) spectrum of the $WSe_2$ thin films. The XRD peaks for all the thin-films could be seen in Fig. 2(a). The measured XRD peaks of $WSe_2$ films match reasonably well with JCPDS data card no 71-0600. One peak was consistently obtained at an angle $27.5^\circ$ which is a consequence of (004) plane. The crystallite sizes were estimated through the Debye- Scherrer formula while strain and dislocation density is calculated using the recipe prescribed in the Ref. \cite{patel2013growth}. Fig. 2(b) shows the variation of strain and dislocation density as a function of $WSe_2$ film thickness. The variation of dislocation density and strain exhibit a similar trend and maximize for the film of thickness $326~nm$. This XRD spectrum distinctly depicts the presence of defects for thicker thin-films which would be investigated later in detail. \\
Raman spectroscopy was employed to ascertain the vibrational modes of the deposited thin $WSe_2$ ﬁlms. Fig. 3(a) shows the Raman spectra for the thin-film samples in the thickness range of 34-326 nm. All the spectra show three characteristic phonon modes. The peak corresponding to $E^{1}_{2g}$ mode is associated with in-plane vibration of $W$ and $Se$ atoms, and the sharp peak corresponds to $A_{1g}$ is related to out-of-plane vibration of $Se$ atoms. The peak corresponding to $B^{1}_{2g}$ mode is essentially due to vibration of $W$ and $Se$ atoms owing to the interlayer interactions \cite{zhao2013lattice,huang2014large,liu2015chemical,sierra2020synthesis}. All the peak intensities and shifts were calculated by fitting the Raman spectra with multiple peaks as shown in fig.3(b) for the 34 nm sample. The inset shows the schematic of displacement of W and Se atoms representing the plane of vibration in Raman active modes. For bulk $WSe_2$ crystal, $E^{1}_{2g}$ and $A_{1g}$ modes are degenerate, yielding single peak as compared to other TMDs materials. Sahin et.al \cite{sahin2013anomalous} demonstrated that when the crystal symmetry broken even by a small amount of uniaxial strain, the degeneracy lifted off and two peaks are clearly observed. We observed that nearly 9-10\% of strain exist for 34 nm samples and increases with higher thickness due to large number of dislocations calculated from XRD, as shown in Fig 2(b).  Such starins are obvious in room temperature sputter deposited thin films. This is the reason that we have observed two distinct $E^{1}_{2g}$ and $A_{1g}$ phonon modes, in all our samples. The variation of Raman-active modes as a function of $WSe_2$ film thickness is shown in Fig. 3(c). It has been observed that the Raman shift for all the vibrational modes reduces for the film of thickness $242~nm$, which can be attributed to the surface defects and that corroborated with the increased surface roughness observed in the FESEM image of Fig. 1(b). The slight change in the Raman shift for a film thickness of $326~nm$ arises because the surface uniformity improves with higher thickness. \\
\begin{figure}
\includegraphics[width=0.35\textwidth]{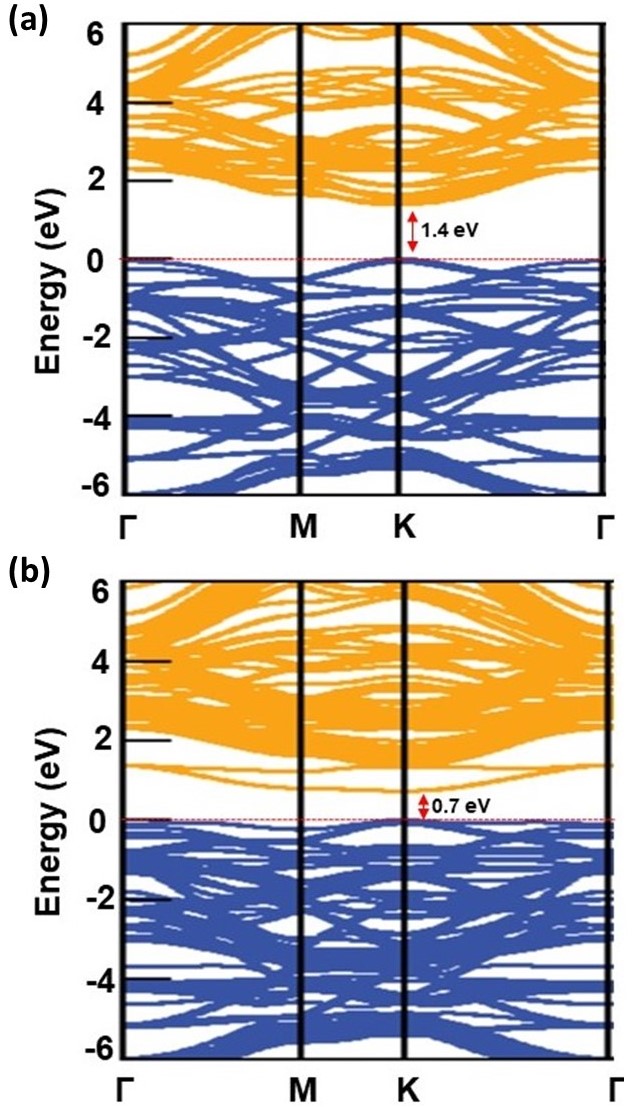}
\caption[bandgap]{Simulated bandstructure of (a) pure $WSe_2$ crystal (b) $WSe_2$ crystal in presence of a defect induced by $Se$ deficiency.}
\end{figure}
The linear absorption spectrum for all the $WSe_2$ thin films are measured using \emph{UV-visible} spectroscopy which is shown in Fig. 4(a). The absorbance tends to reduce at longer wavelengths for all the samples. However, a small increase in absorbance for $WSe_2$ films with thicknesses $\geq~150~nm$, depicts the presence of dual bandgap. From the Tauc fitting expression given by \cite{tauc1966optical,seo2019direct},
\begin{equation}
\alpha h \nu = A(h \nu - E_{g})^n
\end{equation}
where $\alpha$, h$\nu$, A and $E_{g}$ are the linear absorption coefficient, photon energy, a constant and the bandgap respectively, we obtain the bandgap of $WSe_2$ films by extrapolating the linear part of the plot between $(\alpha h \nu)^{1/2}$ and $h\nu$ (see inset of Fig.4(a)). The variation of bandgap(s) as a function of $WSe_2$ film thickness is shown in Fig. 4(b) which distinctly depicts the existence of dual bandgap arising due to the presence of defects in the films. From one perspective, the appearance of mid-gap states (or bands) is attributed to the $Se$ (lighter atom) deficiency in the thin-films grown using RF-sputtering technique. In other words, the deposition method introduces stoichiometric defects (absence of $Se$ atoms) which is distributed randomly throughout the thin-film architecture. \\
In order to gain a deeper insight, we simulated the electronic bandstructure for a $WSe_2$ crystal using density functional theory (DFT) which utilizes the Vienna ab-initio simulation package (VASP) code \cite{kres1}. The interaction
between electrons and ions is introduced by using Projector-Augmented Wave (PAW) method
\cite{bloch}. In these calculations, the structural relaxation is accomplished by deploying generalized gradient approximation (GGA) scheme for adopting the exchange–correlation potential with the Perdew, Burke, and Ernzerhof (PBE) functional \cite{perdew}. A cut-off energy of $350~eV$ is fixed for plane-wave basis set for all the calculations. The Monkhorst–Pack method \cite{monks} are redacted to generate the $5\times 5 \times 5$ $k$-point meshes for achieving the required convergence within $10^{-6}~eV$ per atom. When the forces reached smaller than the value of $0.01$ $eV/\AA$ for each atom, the relaxation is assumed to be completed. The band structure and density of states (DOS) estimations are obtained by considering the geometry which results in most stable configuration and we obtain desirable convergence using the linear tetrahedron method with Bloch corrections \cite{blo}. Figures 5(a) shows the simulated bandstructure for $WSe_2$ which distinctly depicts a bandgap of $1.4~eV$ at the high-symmetry $K$-point in the Brilliouin zone. Also, it is worth mentioning that the energy difference between the conduction band (yellow solid curve) and the valence band (blue solid curve) at other high symmetry points ($\Gamma$ and $M$) are greater than that at the $K$-point which is expected for a hexagonal lattice. At this point, we introduce a $Se$ defect (absence) from a supercell comprising of 16 $W$ atoms. This results in a bandstructure shown in Fig. 5(b) where the defect band is separated from the valence band by about $0.7~eV$ at the $K$-point. Also, it is apparent that the mid-gap defect band is degenerate at the $\Gamma$-point ($\approx~1.4~eV$) and varies between $0.7~eV$ to $0.8~eV$ between $M$-point and $K$-point in the Brilliouin zone. It is worth recalling that the bandgap estimation of $WSe_2$ films using the UV-visible absorption spectrum in Fig. 4(b) which shows the existence of a defect band and it varies between $0.8-1.0~eV$ for film thicknesses $\geq~150~nm$. Within the computational limitation of exactly positioning and identifying the defect sites, the numerically simulated bandstructure provides a plausible explanation for the absorption characteristics and existence of dual bandgap in $WSe_2$ thin films. The presence of the defect band (due to $Se$ deficiency) has a direct impact on the NLO properties of $WSe_2$ films which is discussed below.

We measured the Z-scan OA transmission of $WSe_2$ thin films at an excitation wavelength of $515~nm$ and a peak on-axis intensity of $I_0\approx 3.20 \times10^{11}$ W/cm$^2$ at the focal point. The OA normalized transmittance is shown in Figs. 6(a)-6(f) respectively. It could be observed from the figures that the OA normalized transmittance exhibited a maxima at the focal point of the lens ($z=0$) where the laser intensity is maximum. This is distinct signature of saturable absorption (SA) behaviour. The normalized transmittance were theoretically fitted using the analytical expression given by \cite{sheik1990sensitive,bikorimana2016nonlinear},
\begin{equation}
\Delta T = 1-\frac{\beta I_{0}L_{eff}}{2^{3/2}(x^{2}+1)}
\end{equation}
where $\beta$ is the nonlinear absorption coefficient, $L_{eff}$ is the effective length of the sample which is defined as $L_{eff}$ = (1 - $e^{-\alpha L})/\alpha$, $L$ is the thickness of the films, $\alpha$ is the linear absorption of the film, $x= z/z_0$ where $z_0$ is Rayleigh length for the laser beam. The best theoretical fit to the experimental measurements are shown by solid (black) lines in Fig. 6 which yields a negative value of $\beta$. This is expected from the samples which exhibit a saturable absorption behaviour. The imaginary component of  $\chi^{(3)}$, namely $\chi_{Im}^{(3)}$ could be obtained using \cite{wang2014broadband}
\begin{equation}
\chi_{Im}^{(3)}= \left[\frac{10^{-7}cn^{2}\lambda}{96\pi^2}\right]\beta
\end{equation}
where $c$ is the light speed in vacuum, $n$ is the refractive index of the thin films and $\chi_{Im}^{(3)}$ is expressed in \emph{esu} units. The estimated values of $\beta$ and $\chi_{Im}^{(3)}$ are given in Table-1.
\begin{figure}
\includegraphics[width=0.48\textwidth]{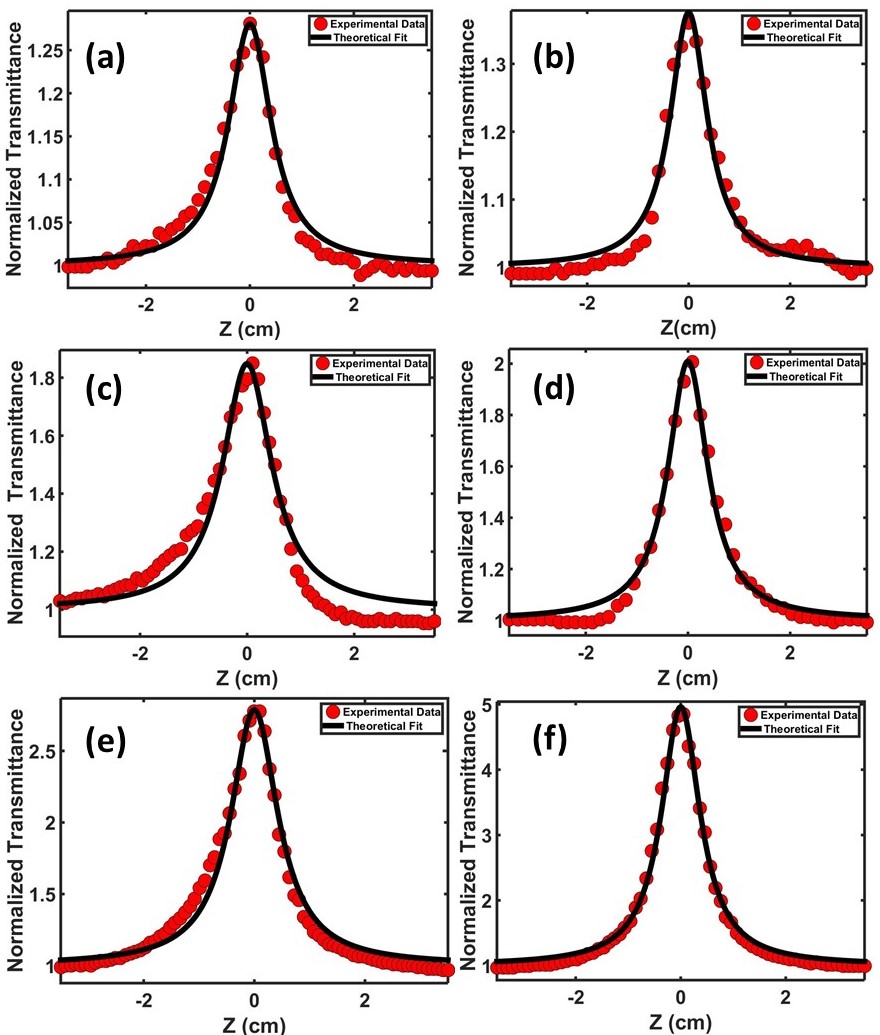}
\caption[Oa]{Open aperture (OA) Z-scan normalized transmittance for $WSe_2$ thin films of thicknesses $34~nm$ (a), $81~nm$ (b), $130~nm$ (c), $150~nm$ (d), $242~nm$ (e) and $326~nm$ (f). Red dots represent the experimentally measured values and black solid curve represent the theoretical fitting (solid lines).}
\end{figure}
\begin{figure}
	\includegraphics[width=0.48 \textwidth]{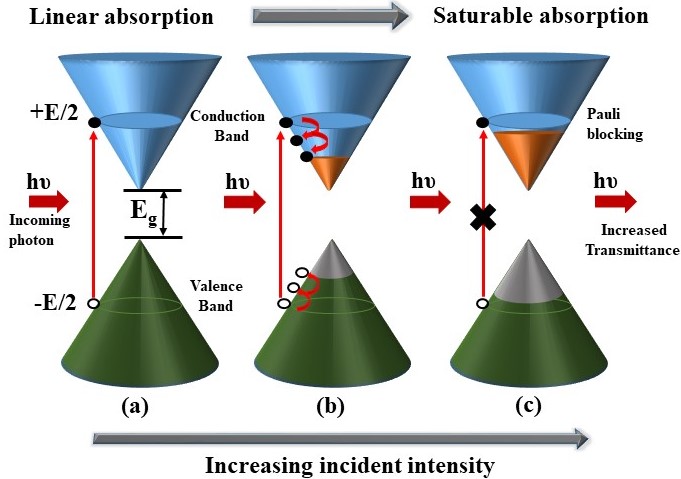}
	\caption[thickness]{A schematic to show saturable absorption mechanism of $WSe_2$ thin films for demonstrating the Pauli-blocking principle. (a) shows the inter-band transition process. (b) shows the process of thermalization of hot electrons and obtaining a thermal equilibrium. (c) shows the filled states near the band edge which restricts the absorption of light.}
\end{figure}
In order to eliminate the impact of linear absorption ($\alpha$) on the nonlinear absorption ($\beta$), we define a figure of merit (FOM) as FOM = $| \chi_{Im}^{(3)}/\alpha |$ which is also mentioned in Table-1. As expected, it could be noted that the linear absorption monotonically increases from $4.25~cm^{-1}$ to $19.07~cm^{-1}$ as a function of increasing $WSe_2$ film thickness. Also, $\beta$ exhibits an approximately $25$-fold increase when film thickness increases from $34~nm$ to $326~nm$. Except a small irregular behaviour for $130~nm$ thin-film, the FOM also exhibits a increasing trend with the film thickness. Consequent upon this observation, it could be safely inferred that the $WSe_2$ thin-films exhibit a SA behaviour at an excitation wavelength of $515~nm$ which tends to increase with film thickness. Although not shown here, the OA Z-scan measurements show a similar SA behaviour for all excitation wavelengths in the visible band of the electromagnetic spectrum. \\
In order to get a deeper insight into the SA characteristics exhibited by $WSe_2$ thin-films, we note that the primary bandgap ($E_g$) for most of the $WSe_2$ films are in $1.2-1.4~eV$ range (see Fig. 4(b)). Therefore, one-photon absorption is expected to be the most dominant absorption process for $515~nm$ wavelength excitation \cite{van1985m}. A schematic viewgraph in Fig. 7(a) shows that the incident wave at $515~nm$ wavelength leads to the excitation of electrons from valence-band (VB) to the  conduction-band (CB) in $WSe_2$ thin films. When the laser intensity is small (at $\vert z\vert >> 0$), the intra-band transitions in CB quickly attains a distribution dictated by Fermi–Dirac statistics for the newly-generated electron-hole pairs. In other words, the electron-hole recombination yields an equilibrium state via intra-band phonon scattering and the electrons (in CB) as well as the holes (in VB) exhibits a distribution as shown in Fig. 7(b). On the other hand, the incident laser intensity is high around the focus ($z = 0$), which leads to significant increase in the number of charge carriers in respective bands. In the thermal equilibrium, the states near the bottom of CB (for electrons) and those near the top of VB (for holes) are completely occupied. This also manifests into a densely populated states (within the band) which are involved in the \emph{two-photon} (nonlinear) absorption process (see Fig. 7(c)). Consequently, the \emph{two-photon} transition probability reduce as a result of non-availability of states in the relevant CB which is alternately termed as \emph{Pauli-blocking effect} \cite{guo2019two,xu2020palladium}. Due to the weak \emph{two-photon} absorption process, the transmission of the films (near the focus) is more than that for regions far-from-focus which could be observed in Fig. 6. In case of $WSe_2$ thin-films, the OA transmission peak at $z = 0$ is higher for thicker films and therefore, it could be inferred that the Pauli-blocking effect is more dominant in thicker films. Further, the bandstructure in Fig. 5(a) depicts that the Pauli-blocking effect is expected to weaken when the \emph{one-photon} (or linear) absorption is small. In order to appreciate this point, we measured OA Z-scan transmittance using ultrashort pulses (of $\approx370~fs$) at excitation wavelengths of $1030~nm$ and $1520~nm$. A representative measurement for two $WSe_2$ thin-films (thickness of $130~nm$ and $242~nm$) are shown in Fig. 8(a) and (b) which depicts a distinct transmission minima at $z = 0$. The solid curves are theoretical fit to experimental measurements (triangular and rectangular dots). Such an OA transmittance  is a signature of reverse saturable absorption (RSA) which could be a consequence of \emph{two-photon} absorption (TPA) in $WSe_2$ thin-films. All the other $WSe_2$ thin-films also exhibit similar RSA signature. It is interesting to observe that the RSA behaviour is stronger at $1520~nm$ as compared to that at $1030~nm$ wavelength for both the $WSe_2$ films shown in Fig. 8. For $WSe_2$ film of thickness $130~nm$, the absence of dual bandgap (or absence of band due $Se$ deficiency) results in a stronger RSA behaviour in comparison with that exhibited by a $242~nm$ $WSe_2$ film which has a distinct dual bandgap. In order to get a deeper insight, we note that the linear absorption in $WSe_2$ thin-films at $1030~nm$ and $1520~nm$ are relatively small which could be observed in Fig. 4(a). This reduces charge-carrier build-up near the band edges which is primarily responsible for Pauli-blocking effect (see Fig. 7(c)). Consequently, the energy eigenstates, which are responsible for TPA process), are likely to have a sparse population density. The TPA probability, therefore, increases which yields OA transmittance signature as shown in Fig. 8(a)-(b). It is apparent from the aforementioned argument that the Pauli-blocking effect would be relatively weak when the linear absorption is small. In consonance with this assertion, we note that the linear absorption of $WSe_2$ is $\alpha = 2.72~cm^{-1}$ at $1520~nm$ excitation wavelength and $\alpha = 4.48~cm^{-1}$ at $1030~nm$ excitation wavelength. Consequently, the OA transmittance minima (at $z = 0$) (in Fig. 8(a)-(b)) for $1520~nm$ excitation wavelength (red-solid curve) has a significantly greater magnitude than that observed in case of $1030~nm$ excitation wavelength. \\
\begin{figure}
	\includegraphics[width=0.48 \textwidth]{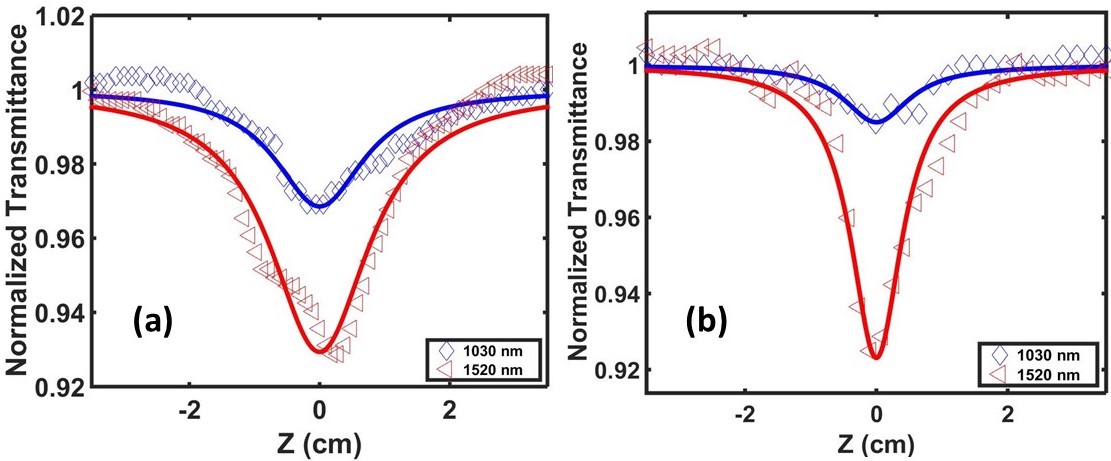}
	\caption[thickness]{OA Z-scan normalized transmittance  curves for (a) $130~nm$ (b) $242~nm$ thick $WSe_2$ films at $1030~nm$ (blue) and $1520~nm$ (red).}
\end{figure}
\begin{table*}
\caption{\label{tab:table1}Nonlinear absorption and nonlinear refraction coefficients of the $WSe_2$  samples at 515 nm excitation wavelength.}
\begin{ruledtabular}
\begin{tabular}{cccccccc}
Sample&$T_0$&$\alpha$&$\beta\times10^{-11}$&$n_{2}\times10^{-16}$&$\chi_{Im}^{(3)}\times10^{-14}$&$\chi_{R}^{(3)}\times10^{-16}$&FOM$\times10^{-15}$\\ 
(nm)&$(\%)$&$(cm^{-1})$&$(cm/W)$&$(cm^2/W)$&(esu)&(esu)&(esu cm)\\ \hline
 34&$65.40$&$4.25$&$-3.01$&$2.81$&$-1.39$&$0.25$&$3.27$\\
 81&$57.76$&$5.49$&$- 4.37$&$3.50$&$-2.13$&$0.56$&$3.88$\\
 130&$43.65$&$8.29$&$- 11.04$&$17.96$&$-2.47$&$1.31$&$2.98$\\
 150&$36.70$&$10.02$&$- 14.64$&$17.98$&$-5.44$&$2.18$&$5.43$\\
 242& $28.27$&$12.63$&$- 28.03$ &$34.79$&$-11.13$&$4.49$&$8.81$\\
 326&$14.84$&$19.07$&$- 76.15$&$59.92$&$-33.28$&$8.53$&$17.45$\\
\end{tabular}
\end{ruledtabular}
\end{table*}
 
  \begin{table*}
\caption{\label{tab:table3}Nonlinear absorption coefficients of the $WSe_2$ thin films at 1030 nm and 1520 nm excitation wavelength.}
\begin{ruledtabular}
\begin{tabular}{ccccc}
Laser&Sample&$\alpha_0$&$\beta\times10^{-11} $&FOM$\times10^{-14}$\\
&(nm)&$(cm^{-1})$&$(cm/W)$&(esu cm)\\\hline
 1030 nm&130&$4.48$&$7.50$&$3.64$\\
 &242&$2.72$&$2.38$&$0.76$\\
 1520 nm&130&$3.24$&$58.94$&$33.29$\\
 & 242&$4.17$&$11.02$&$3.45$\\
\end{tabular}
\end{ruledtabular}
\end{table*}
\begin{figure}
	\includegraphics[width=0.35 \textwidth]{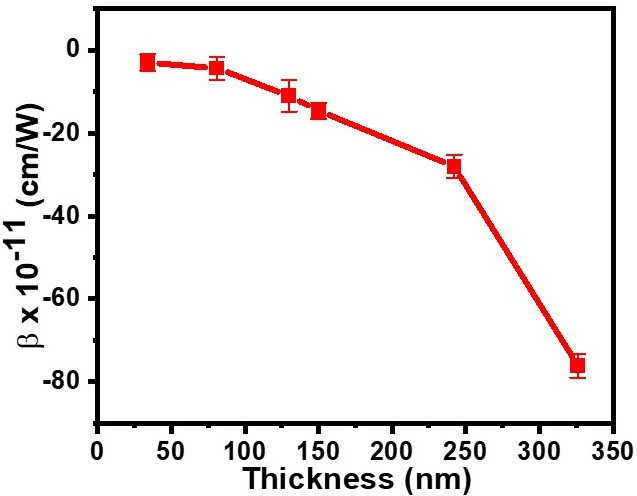}
	\caption[thickness]{Variation of nonlinear absorption (saturable absorption) coefficient $\beta$ as a function of thickness.}
\end{figure}
At this point, it would be interesting to observe the dependence of nonlinear absorption coefficient ($\beta$) on $WSe_2$ film thickness at $515~nm$ excitation wavelength. This is shown in Fig. 9 which depicts an improvement in SA characteristics as a function of thin-film thickness. In order to gain a physical insight, we note that the nonlinear absorption exhibits a direct correlation with bandgap ($E_g$) of the films under investigation. Also, a thicker film is expected to have a higher carrier density owing to denser stacking of $WSe_2$ monolayers. This essentially leads to high occupation of states around the band-edges at high laser intensity (high photon flux), thereby creating a situation for dominant impact of Pauli-blocking. This also manifests into an weaker absorption at high photon-flux (high intensity) which leads to a higher transmission peak at $z~=~0$ in Fig. 6 for thicker $WSe_2$ films. Alternately, this results in stronger saturable absorption features in thicker $WSe_2$ films as depicted in Fig. 9. It could also be observed that the enhancement in SA behaviour increases significantly for $WSe_2$ film thickness $\geq~150~nm$. A natural consequence of denser stacking of $WSe_2$ monolayers manifests in the form of defects whose impact was discussed in the context of dual-bandgap in Fig. 4(b). At $515~nm$ excitation wavelength, the defect band provide a more favorable situation for charge carrier accumulation near the band edges, thereby fostering the possibility of a stronger impact of Pauli-blocking effect and improved SA characteristics. Further the presence defect band also alters the RSA features at $1030~nm$ (($0.82~eV$)) excitation wavelength which could be observed in Fig. 8(a) and (b). The RSA behaviour (blue curve/dots) weakens in the $WSe_2$ film of $242~nm$ thickness as compared to that exhibited by $130~nm$ $WSe_2$ film. This is essentially brought about by a stronger Pauli-blocking effect induced by the presence of defect band. At $1520~nm$ ($0.82~eV$) excitation wavelength, the impact of defect band (in $242~nm$ $WSe_2$ film) on the RSA characteristics weaken essentially due to closer spacing of the defect band and the CB (see Fig. 4(b).

It is apparent that the nonlinear absorption properties of $WSe_2$ thin films exhibit strong spectral sensitivity. In the near-infrared (NIR) spectral band, $WSe_2$ films could play a crucial role in optical power limiting applications as a consequence of prominent intensity dependent absorption characteristics. For example, an optimally thick $WSe_2$ film could be employed as intensity (or amplitude) modulator in the NIR spectral band and therefore, they are highly promising candidates for developing monolithically-integrated miniaturized high-speed ($\sim~GHz$ level) modulators for optical communication. On the other hand, in the visible domain Pauli-blocking induced strong saturable absorption characteristics in $WSe_2$ films allow a deep signal modulation which facilitate desirable shaping of optical pulses. In fact, the depth of modulation for pulse generation as well as shaping is directly proportional to $\vert\beta\vert$  \cite{gomez2018influence} and consequently, $WSe_2$ thin-films exhibiting strong Pauli-blocking induced SA behaviour would serve the purpose very efficiently. It has been demonstrated that stable high-energy nanosecond pulses could be generated by employing a few layer $WSe_2$ sheets (acting as saturable absorber) through passive Q-switching in $Pr^{3+}$-doped ZBLAN fiber laser emitting at $635~nm$ wavelength \cite{wu2018size}. In a similar investigation, B. Chen $et~al.$ have employed an arbitrarily thick $WSe_2$ film as a saturable absorber for generating Q-switched nanosecond pulses in erbium doped fiber laser \cite{chen2016tungsten}. This idea has been extended for generating ultrashort pulses through passive mode-locking technique as well \cite{woodward20152d}. In a recent investigation, a few layer alternating $MoSe_2$ and $WSe_2$ nano-sheets have shown generation of stable ultrashort pulses of a few picoseconds in the NIR band \cite{mao2016erbium}. In fact, the broadband SA characteristics exhibited by $WSe_2$ films give rise to the possibility of developing highly-efficient wavelength-tunable pulse-shaping components in the visible as well as NIR spectrum \cite{yin2017large,liu2018tungsten}. In order to compress pulses to very low time-scales (say $\leq~100~fs$), it is essential to proportionally scale-up the  nonlinear absorption ($\beta$) of $WSe_2$ sheets. This flexibility, in general, is limited in all the aforementioned illustrations which is primarily due to the presence of a few  monolayers of $WSe_2$. The RF-sputtered thin-films, which are experimentally simpler to fabricate, provide a plausible route to tailor $\beta$. As shown here, thicker films exhibit larger $\vert\beta\vert$ and consequently, they could be employed situations requiring deeper modulation.\\
\begin{figure}
	\includegraphics[width=0.48\textwidth]{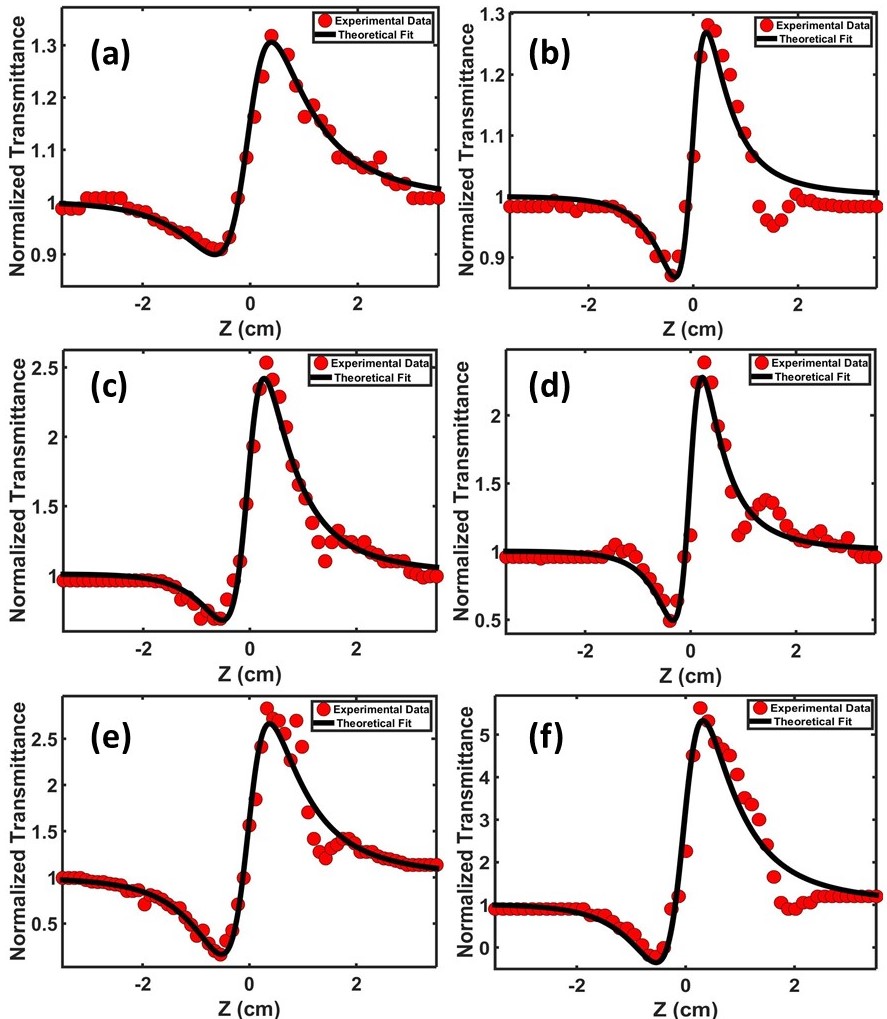}
	\caption[Closed aperture (CA) Z-scan normalized transmittance.]{Closed aperture (CA) Z-scan normalized transmittance at $515~nm$ wavelength for $WSe_2$ thin films of thicknesses $34~nm$ (a) – $326~nm$ (f) respectively.}
	\label{fig:ca Z scan}
\end{figure}
The present investigation also includes the measurements on nonlinear refractive index ($n_2$) using CA Z-scan transmittance. The measured CA normalized transmittance for all the $WSe_2$ thin films at $515~nm$ excitation wavelength is shown in Figs. 10(a)-(f) (red dots). The CA transmittance exhibits a pre-focal minima followed by a post-focal maxima which is a distinct signature of positive $n_2$ values for all the $WSe_2$ thin-films  \cite{boyd2020nonlinear}. However, the asymmetry in the peak-to-valley transmittance is essentially brought about by a strong SA characteristics exhibited by all the thin-films at $515~nm$ excitation wavelength. In order to estimate $n_2$, the normalized CA transmittance curves were fitted using the relation \cite{sheik1989high,sheik1990sensitive,kumar2016phase}
\begin{equation}
\Delta T = 1-\frac{4x\Delta\phi_{0}}{(x^{2}+9) (x^{2}+1)}-\frac{2(x^{2}+3)\Delta\Psi_{0}}{(x^{2}+1)(x^{2}+9)}
\end{equation} 
where $\Delta\phi_{0}= kn_{2}I_{0}L_{eff}$ is the phase change due to nonlinear refraction at the focus, k=$2\pi/\lambda$ is propagation wavevector and $\Delta \Psi_{0}= \beta I_{0}L_{eff}/2 $ is phase change due to nonlinear absorption. The real part of third order nonlinear susceptibility ($\chi^{(3)}$) and $n_2$ exhibit a relation given by \cite{wang2014broadband}
\begin{equation}
\chi_{R}^{(3)}= \left[\frac{n^{2}c}{12\pi^2}\right]n_2
\end{equation}
The values of $n_2$ and $\chi_{R}^{(3)}$ are estimated by fitting (solid black curve in Fig. 10) the experimental measurements using Eq. (4) and represented in Table-1. In case of thinnest film ($34~nm$) thickness, $n_2$ was estimated to be $2.81\times10^{-16}$ cm$^2$/W whereas the thickest film ($326~nm$ thickness) had $n_2 = 59.92\times10^{-16}$ $cm^2/W$ which is approximately a 21-fold increase. 

\section{CONCLUSION} 
In summary, we investigated the nonlinear optical properties of $WSe_2$ thin films using single-beam Z-scan technique at $515~nm$, $1030~nm$ and $1520~nm$ excitation wavelengths. The $WSe_2$ thin films exhibits a SA behaviour and a self-focusing characteristics at $515~nm$ excitation wavelength. The magnitude of nonlinear absorption tends to increase for.thicker $WSe_2$ films. The same $WSe_2$ films exhibit RSA characteristics at both $1030~nm$ and $1520~nm$ wavelengths. The DFT based theoretical modeling provides a plausible route to obtain deeper physical insight into the underlying mechanism for change in the sign of the nonlinear absorption coefficient ($\beta$). In addition, the spectral sensitivity of $\beta$ could be correlated with the electronic bandgap and therefore, it could be tailored by choosing an appropriate film thickness. The $WSe_2$ thin-films of optimum thickness could, therefore, find applications in realizing devices which require a varying degree of optical modulation during operation.
\section{ACKNOWLEDGMENTS}
We thank the Department of Atomic Energy (DAE), Government of India for funding the research work through the project number \emph{RIN-4001}. The authors would like to thank Mrinal K. Sikdar for his help during thin film deposition.

\nocite{*}

\bibliography{apssamp}

\end{document}